# Directed Network of Substorms Using SuperMAG Ground-Based Magnetometer Data

L. Orr[1], S. C. Chapman[1], and J. W. Gjerloev[2,3]

[1]Centre for Fusion, Space and Astrophysics, University of Warwick, Coventry, UK, [2]Applied Physics Laboratory-John Hopkins University, Laurel, MD, USA, [3]Birkeland Centre, University of Bergen, Bergen, Norway**Abstract** We quantify the spatiotemporal evolution of the substorm ionospheric current system utilizing the SuperMAG 100+ magnetometers. We construct dynamical directed networks from this data for the first time. If the canonical cross-correlation between vector magnetic field perturbations observed at two magnetometer stations exceeds a threshold, they form a network connection. The time lag at which canonical cross-correlation is maximal determines the direction of propagation or expansion of the structure captured by the network connection. If spatial correlation reflects ionospheric current patterns, network properties can test different models for the evolving substorm current system. We select 86 isolated substorms based on nightside ground station coverage. We find, and obtain the timings for, a consistent picture in which the classic substorm current wedge forms. A current system is seen premidnight following the substorm current wedge westward expansion. Later, there is a weaker signal of eastward expansion. Finally, there is evidence of substorm-enhanced convection.

**Plain Language Summary** Space weather makes beautiful auroral displays (the northern and southern lights), but with these come large-scale electrical currents in the ionosphere, which generate disturbances of magnetic fields on the ground. These are observed by >100 magnetometer stations on the ground, and the challenge is to extract the important information from these many observations and present it as a few key parameters that indicate how severe the ground impact will be. Networks are now a common analysis tool in societal data, where people are linked based on various social relationships. Other examples of networks include the world wide web, where websites are connected via hyperlinks, or maps, where places are linked via roads. We have constructed networks from the magnetometer observations of space weather events (geomagnetic substorms), where magnetometers are linked if there is significant correlation between the observations. There has been considerable debate as to how the ionospheric pattern evolves during a geomagnetic substorm. We are able to use the networks to resolve some of these controversies.## 1. Introduction

Substorms, their associated current systems, and the corresponding geomagnetic displacements seen at Earth have been the subject of longstanding interest (Pulkkinen, 2015). The fundamental morphology, stages of development, and their timings are well established (Akasofu, 1964). The classic scenario is that of the formation of a substorm current wedge (SCW; McPherron et al., 1973), a rapidly appearing, intense westward electrojet that follows disruption to the cross-tail current system. This corresponds to the DP1 pattern of magnetic perturbations in the nightside auroral zone, which appears in addition to the DP2 geomagnetic counterpart associated with the convective system in the dawn and dusk auroral zones (Nishida, 1968). However, there have been several important variants of this picture. Kamide and Kokubun (1996) proposed a two component auroral electrojet, and Sergeev et al. (2011) argued that their computational wedge model is more consistent with observations, if an additional region two polarity field-aligned current is added to the classic SCW cartoon. Gjerloev and Hoffman (2014) proposed a two-wedge current system, comprised of a bulge and an oval current wedge, in their empirical model of the ionospheric equivalent current system, during an auroral substorm. Recently, it was proposed, by Liu et al. (2018), that there is no large-scale westward electrojet but rather many small, individual segments. These proposed models point to the outstanding question: What is the average substorm current system morphology that we can quantify and resolve uniquely from the full set of available ground-based magnetometer observations? The goal of this paper is to construct a method that quantifies the time-evolving spatial pattern seen across all 100+ magnetometers, in a manner that allows systematic averaging across may substorm events. This will provide a quantitative benchmark to test against model predictions. Key aspects of many of the above models, while being physically distinctive, are qualitative. Our results place these qualitative predictions in direct contact with the observations and can thus drive forward the formation of quantitative hypotheses that will allow these models to be distinguished.

The SuperMAG initiative (Gjerloev, 2012) makes the full set of 100+ ground-based magnetometer observations routinely available, with a standardized coordinate system and a common baseline, supporting both single event and comparative statistical studies. In this form the data are now amenable to analysis methodologies designed to quantify spatiotemporal pattern in sets of multiple, spatially distributed observations. Complex network methodology has recently grown in popularity as a useful mathematical tool and has been used to analyze complex systems from a variety of disciplines ranging from social sciences (Albert & Barabási, 2002; Newman, 2003; Watts & Strogatz, 1998) to geophysical data (Boccaletti et al., 2006; Malik et al., 2012; McGranaghan et al., 2017; Stolbova et al., 2014; Wiedermann et al., 2016). Crucially, unlike other data assimilative methods, including the assimilative mapping of ionospheric electrodynamics, Assimilative mapping of ionospheric electrodynamics (AMIE) (Richmond & Kamide, 1988), network analysis does not introduce spatial correlation. Furthermore, our network analysis does not require any a priori assumptions for variation in ground conductivity since we normalize for this using solely the data to determine the time- and station-dependent network threshold.

Dods et al. (2015) recently demonstrated, on a small set of events, that a network methodology could be applied to the full set of magnetometers for single isolated substorms to yield a characteristic network signature of substorm onset. The networks are time-dependent, hence contain information on the timings of substorm evolution (Dods et al., 2017). Canonical correlation is used to study correlations between multivariate data sets (Reinsel, 2003). If we have two vector time series, canonical correlation analysis will determine the linear combination of the two, which are maximally cross-correlated. The cross-correlation between these linear combinations is the (first) canonical cross-correlation (CCC) component. The key elements of this network analysis are (i) to calculate the CCC of the vector magnetic field time series between each pair of magnetometers and (ii) to apply a station and event specific threshold to this CCC, which is obtained directly from the data. The station pairs that have CCC above the threshold then form a time-varying network.

The analysis of Dods et al. (2015) only examined the *undirected* network (zero-lag CCC). This was sufficient to reveal the initial formation of the SCW at substorm onset but without directional information could not capture the full spatiotemporal evolution of the current system. In this paper we construct the networks based on the (often nonzero) time lags at which the CCC between each pair of stations is maximal, to form the substorm *directed* network, which captures the direction of information propagation between network nodes (magnetometers). Looking across a range of CCC lags captures the full pattern of spatial correlation and how it evolves in time. Nonzero CCC lags indicate the timescale for propagation or expansion of a coherent structure and the sign of the lag gives the direction of propagation or expansion. We construct specific subnetworks to test the hypotheses of different proposed models for how the ionospheric current system evolves. The subnetworks isolate different spatial regions and allow us to test for connections between them. We will focus on spatially well-sampled isolated substorm events and establish network parameters that characterize how the magnetometers collectively respond to the SCW. We have identified 86 events that meet the sampling requirements (this is a subset of the substorm list used in the series of papers by; Gjerloev & Hoffman, 2014, and Gjerloev et al., 2007). We find timings for a pattern in the magnetic field perturbations consistent with the SCW formation at onset, which then expands westward to form a coherent current system in the premidnight sector. There is additional weaker, eastward expansion of the SCW, followed by coherent correlation patterns spanning the entire nightside.

The organization of the paper is as follows. In section 2 we describe the methods and the data used to obtain the directed networks. In section 3 we highlight a case study of one substorm and present a statistical survey 86 events, which reveals how on average the spatial pattern of correlation evolves as the substorm progresses. We conclude in section 4.

## 2. Methods
### 2.1. Constructing the Dynamical Directed Network

For each substorm we first construct the full dynamical directed network before dividing it into subnetworks that flag spatial correlation within and between specific spatial regions. These regions are selected to test different proposed models of substorm current patterns. The term "dynamical" is used here as in the networks literature (Jost, 2007). Our analysis cannot resolve short-range fine structures that are smaller than, or of the order of, the intermagnetometer spacings but can test whether long-range spatially correlated patterns exist. The method for forming a network, at zero lag, is detailed in Dods et al. (2015). The magnetometer stations form the nodes of the network and a given pair of nodes are *connected* if the CCC of their vector magnetic field perturbation time series exceeds an event and station specific threshold, as specified in Dods et al. (2015). In summary, the CCC is calculated over a 128-min running window of the magnetic field perturbations observed by magnetometer pairs. The data are at minute resolution, giving a 128-point CCC for each station pair, every minute. The 128-min sliding window is chosen to give sufficient accuracy in the computed cross-correlation function while also capturing the large-scale spatial and temporal behavior of the SCW. Dods et al. (2017) previously demonstrated using model time series that this window length resolves changes on timescales much shorter than that of the window, specifically capturing onset where there is a sharp ramp in activity in time as the SCW forms. A network is calculated for every minute, and all times, $t$, will refer to the leading edge of the window, that is, the last time point spanned by the window (i.e., a window spanning time interval $[T, T+127]$ will have network properties plotted at time $t = T+127$). Each windowed, three-component vector magnetic field time series is (1) linearly detrended, and (2) the CCC is calculated for each station pair; then (3) if the correlation between magnetometers $i$ and $j$ exceeds the maximum of the two station thresholds, then they are connected and are part of the network. For a network with $M$ active magnetometers, an $M \times M$ adjacency matrix, $\mathbf{A}$, is formed, which has $\mathbf{A}_{ij} = 1$, if $i$ and $j$ are connected, and $\mathbf{A}_{ij} = 0$, otherwise. The station specific threshold for each magnetometer station is determined such that the station will be connected to the network for 5% of the month (28 days) surrounding the event. This ensures that all stations have the same likelihood of being connected to the network, independent of their individual sensitivities to an overhead current perturbation, which in turn depends on the individual instrument characteristics and the local time and season-dependent ground conductivity.

Dods et al. (2015) constructed the network using just the CCC at zero lag. Here, we form the directed network by considering the lag at which the CCC is maximal, $\tau_c$, up to a lag of ±15 min. The value of the CCC value at lag $\tau_c$ is used to determine if the stations are connected (exceeds the threshold) and each connection then also has a direction and timescale of propagation of the observed signal, which is spatially coherent between the two stations. This potentially corresponds to the coherent pattern of time-varying ionospheric currents. The adjacency matrix, $\mathbf{A}$, is not symmetric and the sign of $\tau_{cij}$ determines the signal propagation direction for $\mathbf{A}_{ij}$. If the CCC between magnetometer $i$ and $j$ is above the threshold (they are connected), but with $\tau_c < 0$, we can infer that the signal originates at $j$ and propagates toward $i$, $j \rightarrow i$. If $\tau_c > 0$ the propagation is $i \rightarrow j$.

Gjerloev (2012) found that the probability distributions of differences between SuperMAG baselines and official quiet days rarely exceed 20 nT. For consistency we also exclude magnetometers from the network whose time series of magnetic field perturbations never exceed this noise level.

### 2.2. Data and Event Selection

We analyze vector magnetometer time series at 1-min time resolution for the full set of magnetometer stations available from the SuperMAG database. These data are processed as in Gjerloev (2012), such that the ground magnetic field perturbations are in the same coordinate system and have had a common baseline removed. A set of isolated substorm events, occurring between 1997 and 2000, has been previously identified in Gjerloev et al. (2007). These events have been selected such that (i) they are isolated single events optically and magnetically; (ii) the onset location is spatially defined; (iii) bulge-type auroral events; (iv) there is a single expansion and recovery phase (or the end of the event is at the time of a new expansion); (v) the entire bulge region is in darkness to eliminate any terminator effects; and (vi) they are not during magnetic storms ($|Dst| < 30$ nT) or prolonged magnetic activity. The requirement for darkness creates biases as the events with the majority of the nightside in darkness are in the months around winter solstice. Excluding daylit stations does however avoid large differences in ground conductivity between the stations, which would otherwise dominate the CCC analysis. We also require that activity levels are low for a full window of 128 min before the substorm onset. Together these selection criteria, along with the requirement for a sufficient number of stations in the spatial region around onset (described below), give 86 suitably isolated substorms (listed in the supporting information).

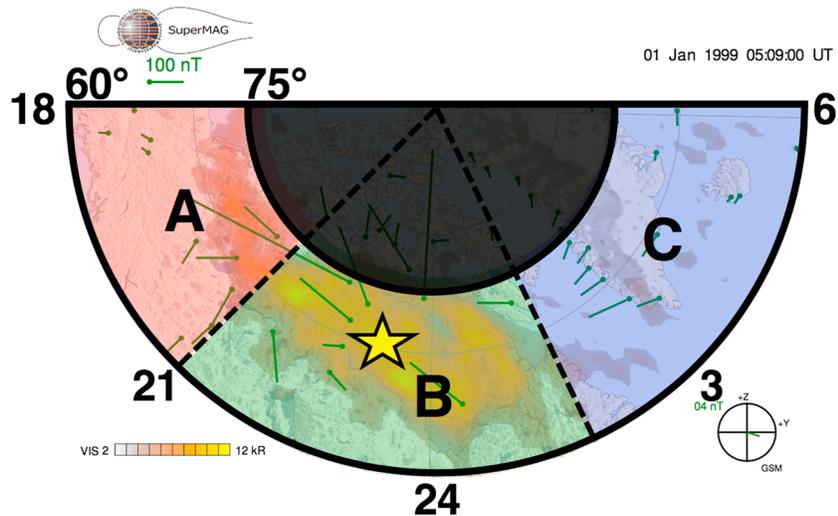

**Figure 1.** SuperMAG polar plot indicating the spatial regions A, B, and C for which we obtain subnetworks. All data are from stations between 60° and 75° magnetic latitude, within the nightside. The local time boundaries between A, B, and C are different for each event and are determined from Polar VIS images; they are separated by the east and west boundaries of the bulge at the time of maximum expansion (dashed lines). The magnetic latitude and local time of onset (again from Polar VIS) for each event are indicated by the yellow star.

### 2.3. Normalization to a Substorm Epoch Time
It is well established that substorms vary in duration (Kullen & Karlsson, 2004; Tanskanen et al., 2002). In order to perform an average across many events, we need to map each event onto a single normalized time base such that, once normalized, all substorms share a common onset time and take the same length of time to evolve from onset to the peak of activity. Following Gjerloev et al. (2007), the observed event time, $t$, is related to the normalized time, $t'$, by

$$t' = \frac{T_E \times (t - t_{onset})}{t_{peak} - t_{onset}}, \qquad (1)$$

where $T_E = 30$ min, approximately the average length of a substorm expansion phase. The onset time is then at $t' = 0$ and the time of peak expansion $t' = 30$. The critical timings for this normalization, $t_{onset}$ and $t_{peak}$, can be unambiguously identified in these isolated substorm events.

### 2.4. Subnetworks for Specific Auroral Spatial Regions
We construct time-varying directed subnetworks that quantify correlation within and between specific spatial regions in the nightside. These spatial regions are selected for each event as shown in Figure 1. The network is constructed using stations located between 60° and 75° magnetic latitude and within the nightside. Gjerloev and Hoffman individually determined the timings and positions of onset and the east and west ends of the bulge portion of the aurora using polar VIS images (Gjerloev et al., 2007). The LT of the bulge edges at the time of maximum expansion ($t' = 30$) has been used to define the boundaries of region B. The study was repeated using the east and west boundaries of the bulge 15 normalized minutes before, and after, the maximum expansion phase ($t' = 15$ and $t' = 45$); results are presented in the supporting information. This gives slight differences, but the overall results and conclusions are unchanged. The SCW is typically 6 hr of local time in extent (Gjerloev et al., 2007), which corresponds to region B; regions A and C are westward and eastward of the SCW, respectively.

We will present a detailed study of the subnetworks for a single event and then will compare it to the average subnetwork behavior seen across all 86 isolated substorms. An event was identified that has ≥7 magnetometer stations in each of regions A, B, and C for the duration of the substorm; this occurs on 1 January 1999 with onset at 04:52 UT. It had a relatively short expansion phase, 17 min, and a thin SCW, extending over 4.1 hr of LT at the time of peak expansion ($t' = 30$). For the averaged study over 86 events we require at least three magnetometers in a spatial region for its subnetwork to be included in the study. For example, a substorm in which there were ≥3 magnetometers in regions A and B, but <3 in C, will contribute to the average subnetworks behavior within A and B but not within subnetwork C. We repeated the entire analysis with the more restrictive criterion of ≥7 magnetometers and found very similar results (see the supporting information). One benefit of using network analysis is that we do not require a spatially uniform grid of magnetometers; that being said, the condition of having ≥3 magnetometers per region gives a mean spatial separation distance (within regions) of ~1,000 km.

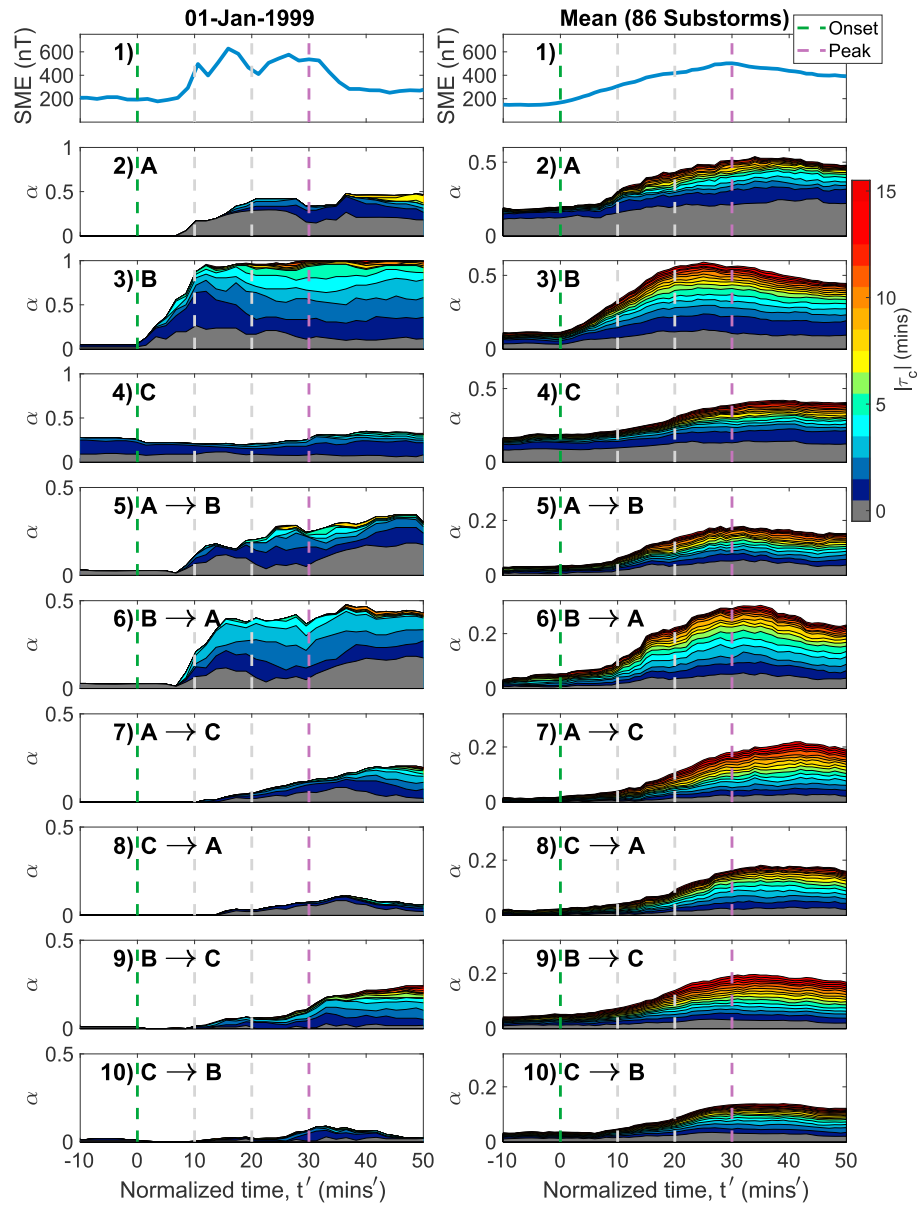

**Figure 2.** The normalized number of connections, $\alpha(t', \tau_c)$, is binned by the lag of maximal canonical cross-correlation, $|\tau_c|$. Each panel stacks vertically (one above the other) $\alpha(t', \tau_c)$ versus normalized time, $t'$, for $|\tau_c| \leq 15$. $|\tau_c|$ is indicated by color (see color bar). Panel 1 plots the SuperMAG electrojet index, SME. Panels 2–10 plot $\alpha$ for connections within and between each of the regions A, B, and C (identified in Figure 1). The left columns plot a single event, whereas the right plots the average of 86 events (containing subnetworks with ≥3 magnetometers per region). Substorm onset (green dashed line) is at $t' = 0$, and the maximum of the expansion phase (purple dashed line) is at $t' = 30$.

The spatial regions A, B, and C are defined such that the subnetworks are always in the same local time relative to the SCW but, as the Earth rotates, the geomagnetic location of the magnetometer stations will vary. This will not affect the properties of the computed network provided regions A, B, and C continue to be well sampled with stations. However, the number of stations within each region can change. We therefore include a normalization to the number of possible connections to define the parameter that we will use to quantify the network, the normalized number of connections:

$$\alpha(t) = \frac{\sum_{i \neq j}^{N(t)} \sum_{j \neq i}^{N(t)} \mathbf{A}_{ij}}{N(t)(N(t) - 1)}, \quad (2)$$

where **A** is the adjacency matrix and $N(t)$ is the number of active magnetometers.

## 3. Results
### 3.1. Observed Timings of Spatial Correlation

We now present (in Figure 2) the directed network for the individual substorm identified above (left column) and the average of all 86 selected substorms (right column). Substorm evolution is not necessarily linear, but the individual substorm is plotted as an example to highlight that the multievent mean is a reasonable average. Having obtained the subnetworks for each region (identified in Figure 1), we have the normalized number of connections, $\alpha(t', \tau_c)$ (equation (2)) within (panels 2–4) or between the regions A, B, and C (panels 5–10). Looking at connected magnetometers within each region provides timings of the emergence of coherent spatial patterns of correlation in the magnetic field perturbations (at ground level), while connections between regions provide information on how these patterns are propagating and/or expanding through out the substorm; any interregion dependencies will also be flagged. If all possible magnetometer connections are present, then $\alpha = 1$. Since the connections between regions (e.g., A→C and C→A) are plotted separately (e.g., panels 7 and 8), then if these were fully connected, the sum over the two plots would be 1. Hence, the range of values for the y axes for connections between regions (panels 5–10) are half the size of that within the regions (panels 2–4).

As the networks are constructed using the time delay/lag at which the CCC between each pair of magnetometer stations is maximal, each connection has an associated signed lag, $\tau_c$. We bin the number of connections ($\alpha$) into ranges of the magnitude of this lag ($\tau_c$). Connections that are at zero lag have no time delay, that is, $\tau_c = 0$ (gray), and connections with an associated direction of propagation/expansion, from one magnetometer to another, have a range of delays, that is, lags from 1–15 min (blue-red). The sign of the lag indicates a direction of propagation or expansion from one magnetometer location to another; this information is combined with the physical geographical locations of the magnetometers to determine if the propagation/expansion is eastward or westward. The connections are separated into different panels for each direction and then binned by the magnitude of the lag. For example, between regions A and C, panel 7 plots the A→C propagation/expansion, eastward, from region A into region C while panel 8 plots C→A propagation/expansion, westward, from region C into region A. Connections with a lag $\tau_c = 0$ (indicated in gray) are plotted on both panels 7 and 8 (A→C and C→A), as they simply indicate instantaneous correlation between regions A and C, which have no associated direction, and thus, by definition the gray bars are identical on the two plots.

Figure 2 stacks the time series of the normalized number of connections, $\alpha(t', \tau_c)$, so that the value of $\alpha$ for each range of $|\tau_c|$ is plotted one above the other for increasing $|\tau_c|$. The stacking is such that each independent $\alpha(t', \tau_c)$ is visible. The envelope is then the total (normalized) number of connections over all lags, that is, all $|\tau_c| \leq 15$. For example, during the individual substorm (left column), we see mostly instantaneous (gray) correlation within A (panel 2) with an additional low level of lagged correlation later in the substorm. On the other hand within B (panel 3), at the time of peak expansion (purple dashed line), the network is made up of ~10% instantaneous correlation, ~80% with $1 \leq |\tau_c| \leq 5$ (fast propagating or expanding), and <10% of connections have $|\tau_c| \geq 6$ (slow expansion or propagation). The plot covers the time interval $-10 \leq t' \leq 50$ normalized minutes where the times of onset, peak expansion, and the 10-min intervals in-between are indicated with vertical dashed lines. The figure presents a summary of time-varying spatial correlation for each subnetwork for the duration of the substorm. The full networks for the individual substorm are plotted in the supporting information. The SuperMAG electrojet index for the individual event, and its multievent average, is plotted in panel 1 of Figure 2. We can see that although the events are on a normalized time base, the multievent average is more smooth and responds less sharply to onset than the individual event.

Prior to onset, the multievent average shows some spatially coherent connections within each of regions A and C (panels 2 and 4). These connections are mostly instantaneous ($|\tau_c| = 0$, gray shading) or with 1-min lag ($|\tau_c| = 1$, dark blue shading). Importantly, these regions are not correlated with each other, so that the number of A→C and C→A connections is small (panels 7 and 8). At onset, in both the individual event and the statistical average, (panel 3) we see that the subnetwork within region B has the most prompt and largest response, that is, increase in spatial correlation. For the individual substorm we see a sharp increase in the number of correlated pairs, beginning at onset ($t' = 0$) and increasing to ~100% (all magnetometers within B are highly correlated) at $t' \sim 15$. Likewise, the multievent average number of correlated magnetometer pairs within the B subnetwork begins to increase at onset, but it is smoother and reaches a peak slightly later than in the individual event; in this case, correlation maximizes with ~60% connectivity at $t' \sim 25$. The January substorm onset is mainly characterized by fast propagating (0- to 6-min lag) connections while in the multievent mean subnetwork ~80% of connections are propagating (nonzero peak lag) throughout. This is consistent with a pattern that is both coherent and propagating and/or expanding. The timings of region B growth are consistent across the majority of substorms observed.

About 10 normalized minutes after onset we can see in panel 6 westward propagation and/or expansion from region B (around onset) into region A (westward of onset), B→A. This coincides with an increase in spatial correlation within region A (panel 2). For both the individual and the multievent average, the B→A time series (panel 6), that is, the relative increase in the number of connections at different lags, resembles that of the network located wholly within region B (panel 3), except that it occurs ~10 normalized minutes later and has about half the magnitude. Within region A (panel 2), ~50% of magnetometers become highly correlated at $20 < t' < 30$. For the individual substorm most of the connections between magnetometers are instantaneous, but for the multievent average $\sim \frac{2}{3}$ of the increase in the number of connections is at nonzero lag. We have found some variation between individual substorms as to how spatial correlation between magnetometers within region A develops from $-10 \leq t' \leq 50$, with some substorms having no obvious response to onset. The A→B (panel 5) propagation and/or expansion develops on similar timescales to B→A (panel 6), but there are significantly fewer connections (20–30% of magnetometers correlated at peak, $t' = 30$) within the multievent average, with $\sim \frac{1}{3}$ of these connections being instantaneous (zero lag, no direction).

The subnetwork for region C (panel 4, east of the SCW) has the smallest response to substorm onset of any region. The January substorm remains moderately connected (~23%) from before onset until long after peak expansion. The multievent average begins to increase at $10 < t' < 20$, and the region is maximally correlated after peak expansion, $t' > 30$ with ~20% of magnetometer pairs being connected; this pattern of correlation is consistent long into the recovery phase. This is consistent with many of the individual substorms showing little/no response to onset. In panel 9 we see that region B becomes correlated with region C with eastward propagation and/or expansion (B→C) ~10–20 normalized minutes after onset, peaking with ~20% magnetometer pairs correlated at $t' > 30$. In panel 10, we can see that for the individual event <10% of magnetometers are correlated from C→B, with this small increase only occurring ~25 normalized minutes after onset. Correlation increases by ~15% for the multievent average between $t' \sim 10$ and $t' \sim 30$. Thus, the response within region C simply tracks that of the propagation or expansion from B → C, and any propagation from C → B occurs subsequently.

Finally, ~10–20 normalized minutes after onset, in panels 7 and 8 we see correlation growing relatively slowly between regions A and C (west and east of the onset location, respectively), reaching a maximum level of connectedness at $t' \sim 40$, long after the time of peak expansion. There is slightly more eastward propagation (A→C, >20% of magnetometers) than westward propagation (C→A, ~10% and ~18% of magnetometer pairs for the individual and multievent average, respectively). Again, the lagged correlation originating in C is very small (and mostly instantaneous) for the individual event.

### 3.2. Interpretation

If we can interpret coherent patterns of spatial correlation across the distributed SuperMAG magnetometers as the emergence of current systems, the above time-dependent network provides an evolution sequence, with timings, for the substorm current system in the nightside. Our analysis then provides a quantitative measure of spatial coherence as well as the timescales on which evolution occurs. By separating the nightside into three regions, we have attempted to isolate the components that have been proposed. To use the terminology of Kamide and Kokubun (1996), we have (A) eastward electrojet; (B) substorm unloading component (SCW), and (C) westward electrojet. Whereas B is associated with the SCW, or DP1 perturbations, A and C can be related to the general magnetospheric convective system, DP2 (Nishida, 1968), which is enhanced during substorm growth and expansion phases (Milan et al., 2017). To summarize the above results, we identify key time ranges, before and after onset: $0 \leq t1' < 10$, $10 \leq t2' < 20$, $20 \leq t3' < 30$ and $t4' \geq 30$ in terms of normalized time, $t'$. Relating these intervals to substorm evolution $t1'$ is following onset, $t2'$ is expansion phase, $t3'$ is near substorm peak, and $t4'$ is the early recovery phase. The timings are as follows:

- Before onset, the premidnight and postmidnight regions A and C each have a relatively weak coherent pattern consistent with convection (DP2); notably, A and C are not coherent with each other.
- In $t1'$ we first see the formation of a SCW (SCW/DP1) around onset (correlation within B), which approaches maximum in $t2'$.
- In $t2'$ there is westward propagation and/or expansion of the SCW west toward the premidnight region (A). We see connections (B→A) and at the same time a signature of a coherent current system within region A (correlation within A). This is shortly followed by weaker correlation from A→B, indicating that the entire A-B system is now correlated. These all approach a maximum at $t3'$.
- A weaker signal of eastward propagation and/or expansion of the SCW toward the postmidnight region starts in $t2'$ and reaches its maximum in $t4'$. We see connections (B→C) and on a similar timescale a signature of a coherent current system within region C (correlation within C), with additional weaker correlation from C→B. The correlation in region C is relatively low.
- The regions eastward and westward of onset (A and C) each have a coherent pattern consistent with enhanced magnetospheric convection (DP2). Later in the substorm there is coherence between regions A and C, beginning well after onset, in $t2'$, and reaching maximum correlation in $t4'$; that is, only after region B has become correlated with all other regions. This can either reflect direct correlation between A and C, or could simply imply that both A and C are correlated with B.

We can then consider what support these results provide for proposed models for substorm current systems, specifically, models with a single westward electrojet segment (Kamide & Kokubun, 1996; McPherron et al., 1973), a westward, and a lower (but still in the auroral zones) latitude eastward electrojet segment (Ritter & Lühr, 2008; Sergeev et al., 2011, 2014); two unconnected westward electrojet segments premidnight and postmidnight (Gjerloev & Hoffman, 2014; Rostoker, 1996); and finally, many small individual segments (Liu et al., 2018). Importantly, any method for quantifying spatial correlation cannot distinguish between direct correlation (here, A→C) and indirect correlation (here, A→B→C); this indirect correlation may enhance the number of A→C connections relative to B→C. Therefore, our results are not inconsistent with multiple separate current systems provided that they are either spatially correlated with each other, or on spatial scales smaller than that of the magnetometer spacing.

The coherent patterns of eastward and westward expansion are in agreement with previous work using synchronous space- and ground-based magnetometers (Nagai, 1982, 1991). However, we have not found definitive support for two, or more, distinct and uncorrelated substorm current systems. A lag, $|\tau_c| > 0$, for B→C connections, implies that C is delayed with respect to B, consistent with a propagation from B to C. Interpreting these results in terms of current components suggests two scenarios for this propagation: (i) a single current segment, which is expanding from B to C, or (ii) a current segment in B and another in C, where the segment in C is correlated with that in B but is developing with some delay. There is no interpretation of our results, which would suggest a scenario where regions B and C are uncorrelated, independent current systems.

If they are associated entirely with general magnetospheric convection (DP2 system), the premidnight and postmidnight (A and C) are directly driven by the solar wind and must enhance on similar timescales, although the magnitudes may differ (Kamide & Kokubun, 1996). We have found that preonset, the regions A and C each have coherent, but relatively small, signatures of correlation with little CCC between them. Postonset, in both the individual event and the average over 86 substorms, the long range east to west (A→C) correlation patterns only emerge after the growth of the SCW (region B). The growth of spatially coherent patterns appears first in B (the SCW, at onset) followed by A (with correlation between B and A) and later, in C. This suggests that following onset, A and C are not solely attributable to enhanced convection, and the presence of contemporaneous B → C and B→ A connections suggests that there may be a combination of contributions from convection enhancement and SCW expansion. Importantly, this does not require that a current segment in A expands or propagates into C.

Finally, if instead of a large-scale SCW there were only many small, uncorrelated, individual segments (Liu et al., 2018), we would not expect to find the long-range correlations (A to C) seen here (also supporting information; Figure 1). Since we calculate CCC on minute resolution time series, each connection in the network is derived from a 128-min time window. Thus, we cannot resolve short-timescale events such as a large number of small wedgelets each associated with a bursty bulk flow in the plasma sheet, which have lifetimes of some 10 min. In addition, we cannot resolve structures that are on smaller spatial scales than the intermagnetometer spacing. If multiple wedgelets are present, their spatial aggregate would give an overall large-scale magnetic amplitude signature mainly at the edges of the region containing the wedgelets, regardless of whether or not the wedglets are spatiotemporally correlated. Here, both spatial and temporal information is used to obtain the cross-correlation so that temporally uncorrelated wedgelets would give no spatially coherent signature of cross-correlation at all, whereas if the same wedgelets were temporally correlated, we would find a signature of spatial cross-correlation.

Potential limitations to the technique include sensitivity to the location of the east and west bulge boundaries, which are static and therefore may not fully represent fast changes in the time-varying current system. There may also be a spatial coarse-graining effect due to the geographic location of the finite number of magnetometers; there are few near the eastward SWC boundary during the January substorm. To test this, we present the same plots for this event, but with the east and west boundaries of the bulge at $t' = 15$ and $t' = 45$ in the supporting information. These show little change from the results shown in Figure 2. Additionally, the detailed network maps of the individual event, for the times represented by the vertical dashed lines in Figure 2, onset-peak, are provided in the supporting information. They highlight the importance of the spatial coverage and geographical locations of the highly correlated magnetometer pairs.

Also, in the analysis and by the organization of data into three regions, A, B, and C, we are quantifying the coherence over these regions. This is over a range in both latitude (60–75°) and local time (typically region B is ~6 hr LT). The westward electrojet around onset (B) may not cover all latitudes, but our analysis technique is mainly addressing the various SCW models, which differ in their local time distribution.

## 4. Conclusions

We used the full set of SuperMAG ground-based magnetometer observations of isolated substorms to quantify the time evolution of patterns of spatial correlation. If the observed pattern of spatial correlation between magnetometer observations captures ionospheric current patterns, then we can directly test different models for substorm ionospheric current systems. We have obtained the first directed networks for isolated substorms. Each connection in the network indicates when the maximum CCC between the vector magnetic field perturbations seen at each pair of magnetometers exceeds an event and station specific threshold. The maximum of the CCC corresponding to each connection in the network can occur at a nonzero time lag. The resulting *directed* network then contains information, not only on the formation of coherent patterns seen by multiple magnetometers but also on the propagation and/or expansion of these spatially coherent structures.

To gain insight on the ionospheric current system during a substorm, we obtained specific time-varying subnetworks from the data that isolate specific physical regions. These regions are west (A), within (B), and east (C) of the bulge boundaries for each substorm (obtained from polar VIS images at the time of peak expansion). We presented both a study of an individual event, which has at least seven magnetometers in each of these regions for the duration of the substorm, as well as the average of the network properties of 86 substorm events. If the observed pattern of spatial correlation between magnetometer observations captures ionospheric current patterns, we find the following sequence of events in terms of key time ranges after onset: $0 \leq t1' < 10$, $10 \leq t2' < 20$, $20 \leq t3' < 30$ and $t4' \geq 30$ ($t'$ is normalized time Gjerloev et al., 2007):

- Preonset, the premidnight and postmidnight regions A and C each have a relatively weak coherent pattern consistent with general magnetospheric convection (DP2) and are not coherent with each other.
- A dominant SCW forming around the onset location (within region B) at the time of onset, $t_1$, which reaches maximum spatial correlation at $t_2$, half way through the expansion phase.
- This is followed by a westward expansion of this SCW (starting at $t_2$, with peak at $t_3$) contemporaneous to and coherent with a current system in the premidnight region (within A).

- An additional weaker eastward expansion of the SCW (starting slowly at $t_2$ with peak at $t_4$). The signal of a self-contained current postmidnight (region C) is relatively weaker and occurs late in the substorm. The enhancement of C is delayed with respect to that of A.
- Following the SCW expansion, A and C are coherent with each other, but at the same time are coherent with the SCW. This is consistent with a combination of convection and expansion of the SCW.

These conclusions are drawn from the averaged network over 86 isolated substorms. Although the overall spatiotemporal timings revealed by this network analysis are reasonably consistent between individual events and the 86 event average for the formation of a SCW around onset (B) and its expansion both east (B→C) and west (B→A), the exact timings of the current system evolution vary. Variability between events could be intrinsic or could relate to the observing conditions, such as differing magnetometer spatial coverage or the static choice of location for region boundaries. Future work will quantify event-by-event variability across multiple events and extend the analysis to multiple, compound events. So far in our analysis we have not utilized the direction of the (vector) maximal CCC. In principle this could resolve the direction of the electojet (eastward/westward).

**Acknowledgments**

We acknowledge use of the SuperMAG ground magnetometer station data from http://supermag.jhuapl.edu/ and thank Rob Barnes for providing us with a hard copy (December 2017). S. C. C acknowledges a Fulbright-Lloyds of London Scholarship, AFOSR Grant FA9550-17-1-0054 and STFCST/P000320/1. We thank the Santa Fe Institute for hosting a visit during which we worked on this research.